\documentclass[prl,a4paper]{revtex4}

\def\ba{\begin{eqnarray}}
\def\ea{\end{eqnarray}}
\def\eaa{\end{array}}
\def\beqar{\begin{array}}
\def\bars{\begin{eqnarray*}}
\def\ears{\end{eqnarray*}}

\def\be{\begin{equation}}
\def\ee{\end{equation}}
\def\f{\frac}
\def\g{\gamma}

\def\ac{\f {\alpha_s N_c}{\pi}}

\def\n{\nonumber \\}

\newcommand{\bt}{\begin{tabular}}
\newcommand{\et}{\end{tabular}}
\newcommand{\bd}{\begin{displaymath}}
\newcommand{\ed}{\end{displaymath}\noindent}
\newcommand{\ec}{\end{center}}
\newcommand{\bc}{\begin{center}}
\newcommand{\ei}{\end{itemize}}
\newcommand{\bi}{\begin{itemize}}
\newcommand{\ii}{\item}

\usepackage{amsmath,amssymb}
\usepackage{epsfig}

\begin{document}
\title{Parametric form of QCD travelling waves}

\author{R. Peschanski}
\email{pesch@spht.saclay.cea.fr}
\affiliation{Service de physique th{\'e}orique, CEA/Saclay,
  91191 Gif-sur-Yvette cedex, France\footnote{%
URA 2306, unit\'e de recherche associ\'ee au CNRS.}}

\begin{abstract}
We derive parametric travelling-wave solutions of non-linear QCD  equations. 
They describe the evolution  towards saturation in the geometric scaling region. 
The method, based on  an expansion in the inverse of the wave velocity, leads to 
a solvable hierarchy of differential equations. A universal parametric form of 
travelling waves emerges from the first two orders of the expansion.
\end{abstract}

\maketitle

{\bf 1.} 
Since the last years, progress is being made in the understanding of  the QCD 
evolution in rapidity towards saturation, i.e. the high-density limit of 
weekly interacting partons probed at high energy. In particular, a 
link  has been established \cite{Munier:2003vc} between travelling-wave 
solutions of non-linear equations and geometric scaling  \cite{Stasto:2000er} 
of the gluon distribution function  ${\cal N}(L,Y),$ where $L\equiv \log k^2$ 
is the transverse momentum variable and $Y$ the rapidity. 

In the mean-field approximation we will consider here,  ${\cal N}$ obeys 
\cite{Balitsky} the nonlinear Balitsky-Kovchegov (BK) evolution equation
\begin{equation}
 \partial_{ Y}{\cal N}=\chi\left(-\partial_L\right){\cal N}- {\cal 
N}^2\ ,
\label{eq:kov}
\end{equation}
where, for convenience,  rapidity $Y$ is measured in units of $\ac$ the strong 
coupling constant, kept fixed. 
\begin{equation}
 \chi(\gamma)=2\psi(1)-\psi(\gamma)-\psi(1\!-\!\gamma)
\label{eq:kernel}
\end{equation}
is the Balitsky Fadin Kuraev Lipatov (BFKL) kernel \cite{Lipatov:1976zz}.

\begin{figure}
\epsfig{file=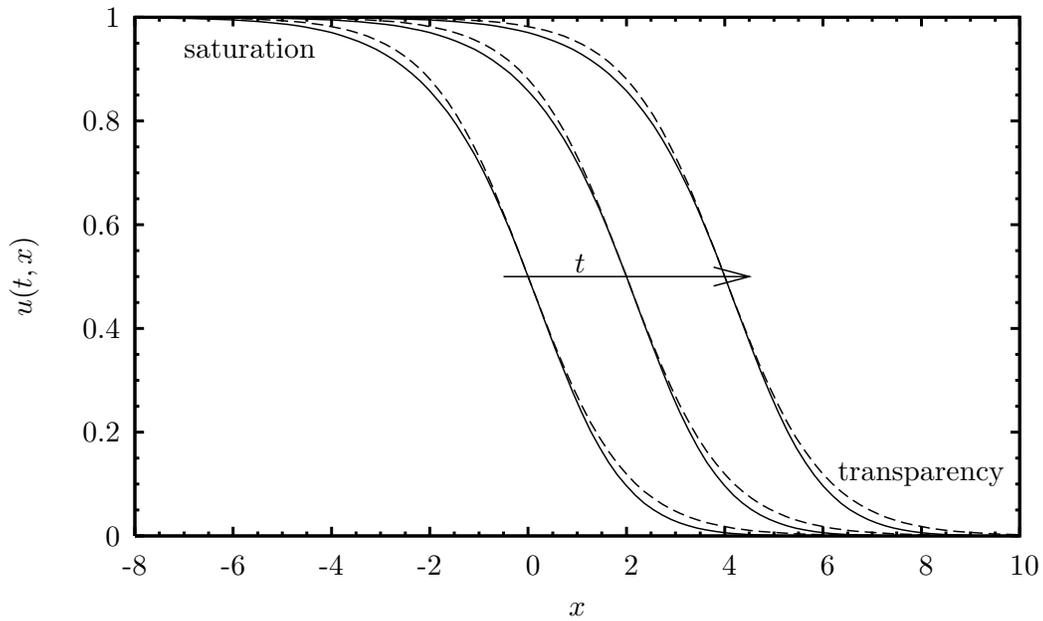,width=14cm}
\caption{\label{fig:fkpp}{\it Universal travelling wave.}
The function $u(x,t)\equiv U(s\!=\!x/c\!-\!t)$ is
represented in the critical case (c=2) for three different times. Continuous 
lines: $U(x/2\!-\!t)$; Dashed lines:  First ${\cal O}(1/c^0)$ term only. The 
wave front connecting the regions
$u=1$ and $u=0$ travels from the left to the right as $t$ increases. That 
illustrates how, in terms of QCD properties,  the ``saturation'' region invades 
the ``transparency'' region. Note that  the first order correction  is already 
small.}
\label{1}
\end{figure}

The identification of Eq.(\ref{eq:kov}), in the diffusive approximation of the 
kernel, with the  Fisher and Kolmogorov-Petrovsky-Piscounov (F-KPP) non-linear 
equation  \cite {KPP} infers \cite{Bramson} the existence of  asymptotic 
travelling wave solutions in rapidity, see Fig.\ref{1}. The mechanism for the 
formation of travelling waves  is  more general \cite{ebert} and  applies also 
\cite{Munier:2003vc} to  the BK equation Eq.(\ref{eq:kov}). Starting with the 
initial condition ${\cal N}(L,Y_0)\sim\exp{(-\g_0L)},$ travelling 
wave solutions ${\cal N}(L\!-\!c\ Y)$  are  formed during the $Y$-evolution with 
the following features  
\cite{ebert}:
\bi
\ii ``pushed front'': $\g_0<\g_c; \ \  c= {\chi(\g)}/{\g},$ with $\g = \g_0$
\ii ``pulled front'': $\g_0\ge\g_c; \ \   c= {{\chi(\g)}/{\g}}, $  with $\g 
=\g_c\ ,$
\ei
where, for the BK equation, $\g_c= .6275$ is the critical  anomalous dimension, 
solution of $\chi(\g_c)/{\g_c}=\chi'(\g_c)=c,$  where $c= 4.883$ is  the 
critical velocity and $\g_0 \!=\!1>\g_c$ (color transparency). For the F-KPP 
equation, the corresponding values are  $\g_c=1$  and  $c= 2.$ These powerful 
results come from a ``locking'' of the solutions of the linear  part of the 
kernel due to the non-linearities. For the ``pulled fronts'', it has universal 
properties, i.e. the locking is independent of the form of non linearities and  
initial conditions. For the  ``pushed fronts'' it is not so.

However, there are  limitations of  these  solutions. They may 
require very large ranges in $Y$ and $L$ to appear. Moreover, while the wave 
structure is clear, the form of the front deduced from the linear kernel does 
not extend to the region where non-linear effects are important. In fact, only 
very few information on the non-linear term has been used. The goal of the 
present paper is to present a different method allowing to analytically derive 
travelling wave solutions taking into account the non-linear terms of the 
equation. 

The values of $\g$ and $c$ coming from the knowledge of asymptotic 
properties  are considered as  input parameters for travelling wave 
solutions of the full non-linear equation, and one uses the fact, already 
noticed in the literature for the F-KPP equation \cite{logan}, that $1/c$ can 
be taken as a small expansion parameter of the full solution.

%%%%%%%%%%%%%%%%%%%%%%%%%%%%%%%%%%%%%%%%%%%%%%%%%%%%%%%%%%

{\bf 2.} 
Let us start with  Eq.(\ref{eq:kov}), where the integro-differential
operator
$\chi\left(-\partial_L\right)$ will  be defined 
with the help of a  limited expansion around the value of $\g$ relevant  for 
the travelling wave solutions, i.e., either $\g_0$ or $\g_c$ depending whether 
we are in a  ``pushed'' or ``pulled'' front situation. 
\begin{equation}
\chi\left(-\partial_L\right)=
\chi(\gamma){\bf 1}+\chi ^\prime(\gamma)(-\partial_L-\gamma{\bf 1})
+{\scriptstyle\frac12}\chi ^{\prime\prime}(\gamma)
(-\partial_L-\gamma{\bf 1})^2
+{\scriptstyle \frac16}\chi ^{(3)}(\gamma)
(-\partial_L-\gamma{\bf 1})^3+\cdots\ .
\label{chiexpansion}
\end{equation}
We shall consider truncations of (\ref{chiexpansion}) in the number of 
derivatives to some value $P.$ We are looking  for a solution in the 
vicinity of the selected anomalous dimension $\g$. For instance the 
diffusive approximation equivalent to the F-KPP equation is obtained with $P=2.$ 
The 
validity of the  approximation will be discussed.

A generic travelling wave (i.e. geometric scaling) solution  has  the form 
${\cal N}(z\equiv bL-a Y),$ with $a,b$ fixed parameters. Inserting it into 
Eq.(\ref{eq:kov}), the partial differential BK equation (\ref{eq:kov}) turns 
into an ordinary differential equation
\be
A_0 {\cal N}-{\cal N}^2+(a-A_1b)\f {d{\cal N}}{dz}+\sum_{p=2}^P (-1)^p A_p b^p\f 
{d^p {\cal N}}{(dz)^p} =0\ 
,
\label{Pwave}
\ee
where
\be
A_p=\sum_{i=0}^{P-p}(-1)^i \ \f {\chi^{i+p}(\g)}{i!}\ \g^i\ .
\label{coeff}
\ee
With suitable redefinitions
\be
 U\equiv \f{{\cal N}}{A_0};\ \ \ \ \ \ c\equiv \f 
a{A_0}-\f{A_1}{\sqrt{A_0A_2}};\ 
\ \ 
\ 
\ \  s\equiv z/c;\ \ \ \ \ \ b\equiv \sqrt{\f {A_0}{A_2}}\ ,
\label{redefinitions} 
\ee
one obtains the equation
\be
U(1-U)+\f {dU}{ds}+ \f 1{c^2} \f {d^2U}{(ds)^2}+\lambda \sum_{p=3}^P\f 
{(-1)^p}{c^p}\  
{A_p}\f {d^p U}{(ds)^p} =0\ ,
\label{PU}
\ee
with $\lambda\equiv \sqrt{A_0 A_2}.$ 

Eq.(\ref{PU}) provides an expansion of the nonlinear differential equation in 
terms of  $1/c^p$ terms with the $1/c$ term missing.  Limiting 
(\ref{PU}) to the  three first terms ($\lambda\to 0$) gives the equation 
that one obtains \cite{logan} for travelling waves in the  F-KPP equation
\begin{equation}
\partial_t u(t,x)=\partial_x^2 u(t,x)+u(t,x)(1-u(t,x))
\label{eq:KPP}
\end{equation}
by inserting the form $u(t,x)\equiv U(s=x/c\!-\!t).$ Hence in the expansion 
around the F-KPP equation, the critical velocity gets  renormalized from its 
original value to $c \sim 2$ by the redefinitions (\ref{redefinitions}).
We see that adding higher derivatives  around $\g$ up to rank $P$ leads to a 
hierarchy of terms in $1/c^p,\ p\ge 3$ with $c\ge 2.$

Taking $1/c$ as the small parameter of an expansion we are looking for an 
iterative  solution $$h(s) = h_0+ \f 1{c^2}h_2+\sum_{p\ge3} \f 
1{c^p}h_p \equiv\f 12 -U(s).$$ Inserting it into  Eq.(\ref{PU}) translates into 
a  hierarchy 
of equations
\ba
h_0'+h_0^2- 1/4 &=&0\n
h_2'+2h_0h_2+h_0''&=&0\n
h_3'+2h_0h_3+\lambda A_3 h_0'''&=&0\n
h_4'+2h_0h_4+h_2^2+h_2''-\lambda A_4  h_0''''&=&0\n
h_5'+2h_0h_5+2h_2h_3+h_3''+\lambda A_3  h_2'''&=&0
\cdots \ ,
\label{Phierarchy}
\ea 
where  we have restricted (for sake of simplicity) the hierarchy of equations 
to  ${\cal O}(1/c^6).$ 
An iterative solution can easily be found with  initial conditions being 
appropriately chosen, i.e. $h_0(\pm \infty) = \pm \f 12$ and 
$h_{i\ne 0}(\pm \infty) = h_{i}(0)=0$  for higher orders. Note that if $h(s)$ is 
solution, also $h(s+s_0)$ is solution.

The system is iteratively but fully solvable. One first  solve the  only 
non-linear equation of the hierarchy, obtaining $h_0={\scriptstyle \f 12} {\rm 
th}(\f s{\scriptstyle 2})$. Using the property 
\be
\f {d}{ds}h_n(s)+2h_0 h_n(s) \equiv \f 1{ch^2(s/2)}\ \f {d}{ds}\left[ch^2(s/2)\ 
h_n(s)\right]\ ,
\label{iteration}
\ee
all the other linear equations   reduce to simple integrations of functions 
defined recursively from the  hierarchy.
 
The solution of the first three terms of the expansion is 
\begin{equation}
U(s) = \f 1{1\!+\!e^{s}}\!-\!\f 1{c^2}\ \f 
{e^{s}}{\left(1\!+\!e^{s}\right)^2}\  
\log \f {\left(1\!+\!e^{s}\right)^2}{4e^{s}}\! 
+\!\f 1{c^3}\ \lambda A_3\ \f {e^{s}}{\left(1\!+\!e^{s}\right)^2}\ 
 \left(3 \f {\left(1\!-\!e^{s}\right)}{\left(1\!+\!e^{s}\right)}\!+\!s\right)+ 
{\cal O}(\f 1{c^4})\ .
\label{eq:Psolution}
\end{equation}

This can be compared with the hierarchy of equations obtained for the F-KPP 
equation (\ref{eq:KPP})
\ba
h_0'+h_0^2- 1/4 &=&0\n
h_2'+2h_0h_2+h_0''&=&0\n
h_4'+h_2^2+2h_0h_4+h_2''&=&0\cdots \ ,
\label{hierarchy}
\ea
where the expansion parameter is now $1/c^2.$
The solution of the first three equations \cite{logan} 
of (\ref{hierarchy}) is 
\begin{equation}
U(s) = \f 1{1\!+\!e^{s}}\!-\!\f 1{c^2}\ \f 
{e^{s}}{\left(1\!+\!e^{s}\right)^2}\  
\log \f {\left(1\!+\!e^{s}\right)^2}{4e^{s}}\! +\!\f 1{c^4}\left\{ \f 
{e^{s}(e^{s}\!-\!1)}{\left(1\!+\!e^{s}\right)^3}\  
\left({\scriptstyle \f 12}\log^2 \f {\left(1\!+\!e^{s}\right)^2}{4e^{s}}\! -\!
\log \f {\left(1\!+\!e^{s}\right)^2}{4e^{s}}\!+\!3\right) +\f 
{se^s}{(e^{s}\!+\!1)^2}\right\}+ {\cal O}(\f 1{c^6})\ .
\label{eq:solution}
\end{equation}

The comparison of the sets of equations (\ref{Phierarchy}) and (\ref{hierarchy}) 
and their 
solutions  (\ref{eq:Psolution}) and (\ref{eq:solution}) calls for comments.

i) The two first terms of the expansion in the small parameter of order $1$ and 
$1/c^2$ are the same, and then give an ``universal parametric form'' to the 
travelling wave solution, see Fig.\ref{1}. The form is unique while the 
parameters values may change  by a redefinition (\ref{coeff}) of the 
coefficients  $A_i.$  

ii) The non-universal dependence on the kernel appears at order $1/c^3.$ It can 
stay small provided the coefficient $A_3$ and the followers remain bounded. 
This implies to keep the expansion of the kernel limited to the region around 
the chosen value of $\g$ and in particular in the vicinity of $\g_c$ in the 
critical case $c=2.$ 

iii) The truncation of the expansion of the kernel is  required for the BK 
equation equipped with the BFKL kernel. Indeed, if considering the full 
analytic expansion to define the coefficients $A_i$ (i.e. 
$P\to \infty$), they are infinite since they correspond to the expansion of the 
BFKL kernel near $\g=0$ where it possesses a pole singularity. We  interpret 
this phenomenon as a breaking of the travelling wave property in the deep 
saturation region of the BK equation. In fact, both in the deep saturation 
region 
($u\sim 1$ in Fig.(\ref{1})) and in the leading edge of the wave ($u\sim 0$) we 
expect the exact scaling properties not to be valid \cite{ebert}. Scaling is 
expected in between, that is in the ``interior'' of the wave.

%%%%s%%%%%%%%%%%%%%%%%%%%%%%%%%%%%%%%%%%%%%%%%%%%%

{\bf 3.}
Using  the redefinitions   (\ref{redefinitions}) , and limiting ourselves  to 
the 
universal part, we obtain
\be
{\cal N}\propto \f 1{1\!+\!\left[\f {k^2}{Q^2_s(Y)}\right]^{\mu_1}}\!-\!\f 
1{c^2}\ \f 
{\left[\f{k^2}{Q^2_s(Y)}\right]^{\mu_1}}
{\left(1\!+\!\left[\f{k^2}{Q^2_s(Y)}\right]
^{\mu_1}\right)^2}\  
\log \f {\left(1\!+\!\left[\f {k^2}{Q^2_s(Y)}\right]^{\mu_1}\right)^2}
{4\left[{k^2}{Q^2_s(Y)}\right]^{\mu_1}} + {\cal O}\left(1/c^3\right)\ ,
\label{result}
\ee
where
\be
Q^2_s(Y) = \exp \left(\mu_2 Y\right)\ ,
\label{scale}
\ee
plays the role of the saturation scale. In (\ref{result},\ref{scale}), we have 
defined $\mu_1 \equiv \f {\sqrt{A_0/A_2}}c,$ 
$\mu_2 \equiv \left(A_1+c\left[\f {A_2}{A_0}\right]\right).$ 
The solution (\ref{result}) only depends on  the  function 
\be
e^s \equiv \left[\f {k^2}{Q^2_s(Y)}\right]^{\mu_1} = \left[\f 
{W^2}{W_0^2(k^2)}\right]^{\mu_2}\ ,
\label{forms}
\ee
where one writes $W^2=e^Y,\ W_0=k^{-\mu_1/\mu_2}.$ 
Hence $e^s$ has the typical structure of  a forward Regge amplitude. While the 
first form exhibits the geometric scaling structure, the second one suggest an   
interpretation of  (\ref{result}) as an unitarization of the  amplitude 
(\ref{forms}).

Hence, we have found the possibility of assessing a specific 
parametric form to the geometric scaling solutions of the BK equation, through a 
redefinition of the parameters $\mu_1,\mu_2.$  The form   
 (\ref{result}), with the possibility of extending the parametrization to higher 
orders by a solution of the hierarchy (\ref{Phierarchy}), may help establishing 
a link between the theory and the phenomenological observation of geometric 
scaling \cite{Stasto:2000er}. This is the subject of a subsequent work 
\cite{us}.

On a more theoretical level, it seems feasible to extend the method to  QCD 
evolution equations beyond the ``mean field''  BK equation  (\ref{eq:kov}). In 
particular the extension to the stochastic versions of  QCD evolution equations 
\cite{stochastic} could  be treated \cite{us} by averaging the parametric form 
over the distribution of velocities $c$ generated by the noise. 

In conclusion, the fruitful properties of travelling-wave solutions of 
non-linear 
evolution equations, both through the asymptotic limit and  through  their 
expansion in terms of the inverse velocity give significant tools to deepen our 
understanding of QCD in the high density limit and emphasize once more the 
interesting connections with statistical physics.

\vspace{.5cm}

\begin{acknowledgments}
R.P. wishes to thank C. Marquet and G. Soyez  for help and fruitful discussions.
\end{acknowledgments}

%\bibliography{kov}
%\bibliographystyle{h-physrev3}

\end{document}